# Telecom Compatibility Validation of Quantum Key Distribution Co-existing with 112 Gbps/λ/core Data Transmission in Non-Trench and Trench-Assistant Multicore Fibers

R. Lin[1,3], A. Udalcovs[2], O. Ozolins[2], X. Pang[1,2], L. Gan[3], L. Shen[3], M. Tang[3], S. Fu[3], S. Popov[1], C. Yang[4], W. Tong[4], D. Liu[3], T. Ferreira da Silva[5], G. B. Xavier[6,7]*, and J. Chen[1]**

(1) KTH Royal Institute of Technology, Electrum 229, Kista, Sweden, **jiajiac@kth.se
(2) Networking and Transmission Laboratory, RISE Acreo AB, Kista, Sweden
(3) Huazhong University of Science and Technology, Wuhan, China
(4) Yangtze Optical fiber and Cable Joint Stock Limited Company (YOFC), R&D Center, Wuhan, China
(5) National Institute of Metrology, Quality and Technology, Optical Metrology Division, 25250-020, Duque de Caxias RJ, Brazil
(6) Institutionen för Systemteknik, Linköpings Universitet, Linköping, Sweden, *guilherme.b.xavier@liu.se
(7) Departamento de Ingeniería Eléctrica, Universidad de Concepción,160-C Concepción, Chile

**Abstract** *We experimentally characterize photon leakage from 112Gbps data channels in both non-trench and trench-assistant 7-core fibers, demonstrating telecom compatibility for QKD co-existing with high-speed data transmission when a proper core/wavelength allocation is carried out.*

## Introduction

Quantum key distribution (QKD) provides an important alternative to classical encryption algorithms, which are vulnerable to eavesdroppers equipped with quantum computers[1]. The absolute security of QKD depends on handling extremely faint optical pulses. This contrasts with traditional telecom systems that may induce noise onto quantum channels if high-speed data transmission and QKD are co-existing in the same fiber[2].

Spatial division multiplexing (SDM) has emerged as a key technology to keep up with the ever-increasing bandwidth density[3]. In recent demonstrations of QKD over SDM[4-5], the employed multi-core fibers (MCFs) have ultra-low crosstalk (XT) between adjacent cores, showing a high potential to accommodate data communication and QKD in the same fiber. Such high isolation is typically provided by a "trench" in the refractive index profile, therefore this type of MCF is often referred to as trench-assisted MCFs (TA-MCF). Non-trench MCF (NT-MCF)[6], on the other hand, can potentially improve spatial density and simplify the integration of fiber and MUX/DEMUX by removing the trenches[6]. Even though a residual coupling effect is present, the requirement of the prohibitively complex digital signal processing (DSP) can still be avoided. However, whether such coupling effect degrades telecom compatibility in terms of supporting simultaneous QKD and high-speed data transmission over MCFs is still question marked. In this paper we experimentally characterize the penalty for QKD transmission over both TA-MCF and NT-MCF when cores are carrying 112 Gbps pulse amplitude modulation (PAM4) signal per wavelength (λ) data transmission. Based on the measurement results, we verify the telecom compatibility of QKD in TA- and NT-MCFs, leading to a crucial step towards integrating QKD into SDM transmission systems with high bandwidth density.

## Operation principle

We first record the photon leakage from data transmission channels located in different wavelengths and in the cores other than quantum core using NT- and TA-MCFs. Then the measured leakage is used to calculate the lower bound for the secret key rate ($R$), which is an estimate of the key length shared between Alice (the transmitter party) and Bob (the receiver party) after all post-processing[1]. It is expressed as $R \geq Q_1[1 - H_2(e_1)] - Qf(E)H_2(E)$, where $Q_1$ and $Q$ are the single-photon and all photon-number states gain[7] (i.e., probability of photo generation by Alice, and photo detection by Bob), $e_1$ and $E$ are the single-photon and overall quantum bit error rate (QBER), $H_2(x)$ is the binary Shannon entropy of variable $x$, and $f(E)$ is the inefficiency of the error correcting procedure, typically assumed to be 1.22[7]. The noise due to the total leakage is added to the detector dark count probability (which is a part of the terms $Q$, $Q_1$, $E$ and $e_1$), composing the probability of detection by Bob when Alice sends vacuum state, considering the infinite decoy states technique[7]. We then estimate the key rate as a function of transmission distance, taking into account the leakage corrected for fiber distance.

## Experimental setup

In the experiment, we test a 2.5 km long homogenous 7-core TA-MCF (Fig. 1a) and NT-MCF (Fig. 1b), both with a center core and six side cores. The average core pitch and the cladding diameter of the NT-MCF are 41.1 um and 150.0 um. The refractive index (RI) of each core and the cladding are 1.4639 and 1.4591, respectively[6]. The measured XT for the 2.5 km NT-MCF is about −45 dB, while the measured XT for the 2.5 km TA-MCF is lower at ~ −65 dB. Similar attenuation and chromatic dispersion can be found in both types of fibers, which are 0.2 dB/km/core and 16 ps/nm/km. We define the quantum channel at 1550nm in the center core, representing the worst case for quantum channel in terms of inter-core crosstalk. Two distinct layouts are used in the measurements: 1) Only the side cores contain data channels ("*side cores*" configuration); 2) Center core + side cores are populated with data channels, so-called "*all cores*" configuration.

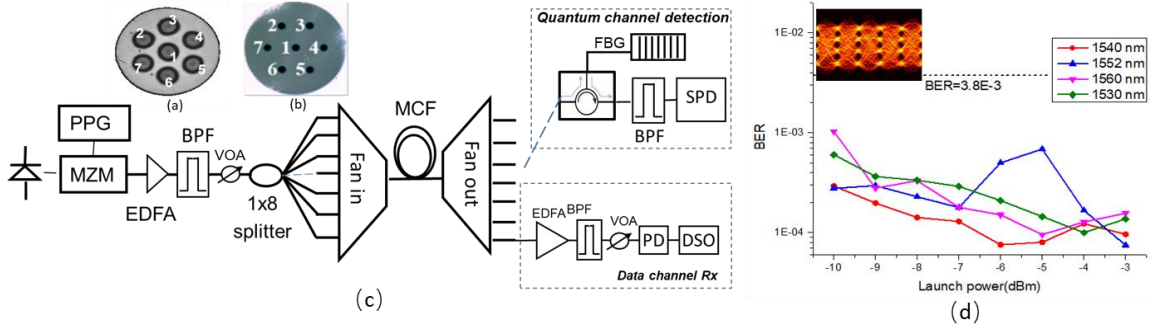

**Fig. 1:** Cross-section for the (a) TA-MCF and (b) NT-MCF; (c) experimental setup; and (d) BER performance as a function of received optical power (RoP) under different launched powers. PPG: pulse pattern generator, VOA: variable optical attenuator, BPF: band-pass optical filter, FBG: fiber Bragg grating, MCF: multi-core fiber, SPD: single-photon detector.

The overall experimental setup is shown in Fig. 1c. A tunable distributed feedback laser operating at C+L band is used as the data channel transmitter. A 56 GBaud PAM4 signal is generated by multiplexing four 28 Gbps non-return-to-zero (NRZ) streams using an Anritsu DAC G0374. The output electrical signal is amplified to drive a $LiNbO_3$-based Mach-Zehnder modulator (MZM) biased in the linear regime. The output optical signal is amplified by an Erbium-doped fiber amplifier (EDFA) followed by a tunable bandpass filter (BPF) with ~30 dB rejection ratio to eliminate the amplified spontaneous emission (ASE) noise, and a variable optical attenuator (VOA) for adjusting the launched power into each core. The optical signal is divided into 7 branches using a 1:8 splitter, temporally decorrelated, and then coupled into the MCF via a fan-in (FI) module. After propagation, the spatial channels are demultiplexed by a fan-out (FO) module so that the signal from each core is connected to single-mode fiber (SMF) pigtails. The high-speed data channel is detected by a receiver consisting on a pre-amplifier with a fixed output power, a BPF and a photodetector (3 dB bandwidth of ~33 GHz) and processed off-line. Another VOA is used to adjust the feed-in-receiver photodetector (PD) power for bit error rate (BER) evaluation. A decision-feedback equalizer (DFE) with 43-feed-forward taps and 12-feedback taps is used to compensate the transmission impairment.

At the reception of the quantum channel, the center core's output is first reflected by a fiber Bragg grating (FBG), centered at 1550 nm and then goes through a BPF (in total ~55 dB extinction ratio), where out-of-band photons from the quantum channel are filtered. The output of the BPF is directly connected to a single-photon detector (IdQuantique id 210), working in gated mode with 1 MHz repetition rate, 10 μs deadtime, 10% overall detection efficiency, and an effective 1-ns wide gate width. Dark count probability is $1.3 \times 10^{-5}$ per gate.

**Results and discussion**

The BER performance as a function of the launch power for data channel is shown in Fig.1d. Here, the worst core in terms of attenuation is selected to obtain the lower bound on the data transmission performance when all cores are loaded. When the launched power to each core is swept from -3 dBm to -10 dBm, the 112 Gbps PAM4 signals can always reach a BER much lower than $3.8 \times 10^{-3}$ (i.e., hard-decision-forward-error-coding HD-FEC limit), across the whole spectrum (1530 nm to 1560 nm). With feed-in PD power of 0 dBm, data transmission beyond 100 Gbps can be achieved.

The detection rate of leaked photons to the quantum channel due to simultaneous data transmission in both "side cores" and "all cores" cases, for a different number of wavelengths in the NT-MCF is shown in Fig. 2a. With a launch power of -4 dBm and lower using wavelength not close to the quantum channel, the photon detection rates in the quantum channel are less than 100Hz regardless of whether the center core is used for data transmission or not. However, when 1552nm (close to the quantum channel) is assigned to data transmission, the detection rate is much higher. A similar measurement is carried out using 1552nm to deliver data in "side cores" configuration for TA-MCF, showing a photon detection rate very close to the detector's dark count limit at ~13 Hz. Compared to TA-MCF a higher inter-core crosstalk in NT-MCF makes a larger number of photons passing through the detection filter.

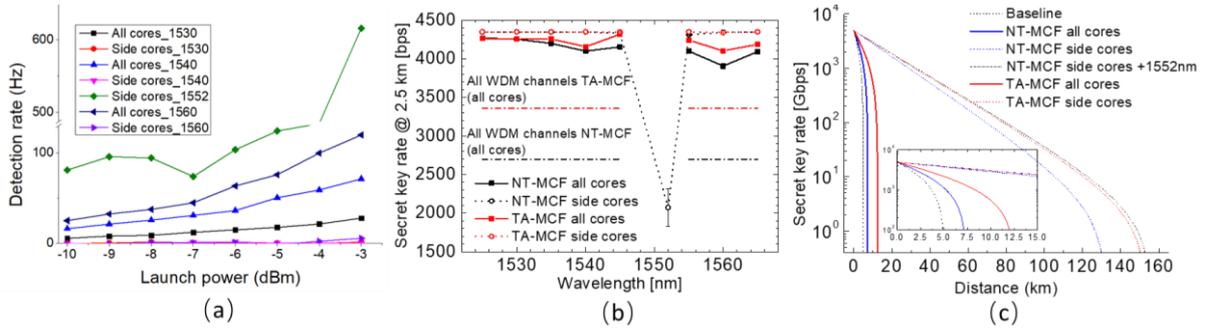

**Fig. 2:** (a) Detection rate of photon leakage for NT-MCF as a function of launch power with different wavelengths assigned to data transmission for "*side cores*" and "*all cores*" configuration; (b) secret key rate at 2.5 km MCF fiber as a function of the wavelength assigned for data transmission. Continuous dashed lines show the value if all measured wavelengths, except 1552 nm, are used for data transmission simultaneously in "all cores"; (c) secret key rate as a function of distance for SMF and both types of MCFs.

Based on the photon leakage measurement, the secret key rate $R$ following the 2.5 km MCF is estimated and plotted in Fig. 2b, with different wavelengths assigned to data transmission for both types of MCFs. With a repetition rate of 1MHz in QKD system, ~4.4kbps secret key rate can be achieved for both fibers in the side cores configuration. A negligible difference on the key generation rate can be observed between the deployment of the two types of MCFs except 1552nm, even though a higher inter-core crosstalk is in the NT-MCF. The exception is when 1552 nm is used as the data channel, key generation can be realized at a considerably lower key rate in NT-MCF compared to that in TA-MCF. When the center core is additionally deployed for data transmission, i.e., "*all cores*" configuration, the quantum channel is affected by the inter-wavelength crosstalk, and a penalty on secret key rate can be seen compared to "side cores" configuration. We also include the expected secret key rate for "all cores" configuration when all the 8 measured wavelengths except 1552 nm are employed simultaneously, where an obvious penalty on key rate can be observed in both types of MCFs.

In Fig. 2c we plot the estimated lower bound of secret key rate as a function of transmission distance. For comparison purpose, we reproduce the results from[8], where SMF is used as the baseline. In the "*side cores*" case, NT-MCF still can perform close to TA-MCF and baseline without assigning 1552nm for data transmission, whereas a significant drop in distance (<5km) if the wavelength close to quantum channel is assigned to data delivery in NT-MCF. In the "all cores" configuration, the secret key rate quickly falls off in both types of MCFs.

## Conclusions
We experimentally validate the possibility of QKD co-existing with 112 Gbps/λ/core in two types of MCFs by evaluating the performance of data channels and its impact on the quantum channel. These results demonstrate that core/wavelength allocation for data transmission is crucial for QKD telecom compatibility in MCF based optical networks. For NT-MCF that has a great potential to improve bandwidth density, the wavelength close to quantum channel is recommended to be avoided in any cores, while for both types of MCFs data transmission carried in the core assigned to the quantum channel could dramatically affect the quality of QKD.

## Acknowledgements
This work was supported by Fondecyt 1150101, Ceniit–Linköping University, Swedish Research Council, the Swedish Foundation for Strategic Research (SSF), the Göran Gustafsson Foundation, the Swedish ICT TNG, the Celtic-Plus sub-project SENDATE-EXTEND & SENDATE FICUS funded by Vinnova, and the National Natural Science Foundation of China. H2020 NEWMAN Project (#752826). The Knut and Alice Wallenberg is acknowledged for equipment funding and Tektronix for equipment loan.